\documentclass[aps,pre,showpacs,onecolumn]{revtex4-1}

\usepackage{amssymb}
\usepackage{amsmath}
\usepackage{graphicx}
\usepackage{color}
\usepackage[normalem]{ulem}
\begin{document}

\title{Stability of Neel skyrmions in ultra-thin nanodots considering Dzyaloshinskii-Moriya and dipolar interactions}

\author{N Vidal-Silva$^1$, A Riveros$^1$ and J Escrig$^{1,2}$}

\address{$^1$Departamento de F\'isica, Universidad de Santiago de Chile, Avda. Ecuador 3493, 9170124, Santiago, Chile.}
\address{$^2$Center for the Development of Nanoscience and Nanotechnology (CEDENNA), Avda. Ecuador 3493, 9170124 Santiago, Chile}

%\ead{alejandro.riveros@usach.cl}
%\vspace{10pt}
%\begin{indented}
%\item[]January 2017
%\end{indented}

\begin{abstract}
An analytical expression for the energy of N\'eel skyrmions in ultra-thin nanodots considering exchange, uniaxial anisotropy, Dzyaloshinskii-Moriya, and dipolar contributions has been obtained. In particular, we have proposed for the N\'eel skyrmion, a general ansatz for the component of the magnetization perpendicular to the dot, given by $m_z(r) = [1-(r/R_s)^n]/[1 + (r/R_s)^n]$, where $R_s$ is the radius of the skyrmion and $n$ is an integer and even number. As proof of concept, we calculate the energy of a N\'eel skyrmion in an ultra-thin Co/Pt dot, and we find that the dipolar contribution cannot be neglected and that both Dzyaloshinskii-Moriya interaction and anisotropy play an important role to stabilize the skyrmion. Additionally, we have obtained a good agreement between our analytical calculations and previously published micromagnetic simulations for $n = 10$. For this reliable value of $n$, we have obtained that for a Dzyaloshinski Moriya constant $D = 5.5 \, (mJ/m^2)$, it is possible to stabilize a N\'eel skyrmion for $K_u$ in the range, $0.4 \, (MJ/m^3)< K_u <1.3 \, (MJ/m^3)$, whereas for $K_u = 0.8 \, (MJ/m^3)$, the skyrmion stabilizes for $5.0 \, (mJ/m^2) < D <6.0 \,  (mJ/m^2) $. Thus, this analytical equation can be widely used to predict stability ranges for the N\'eel skyrmion in spintronic devices.
\end{abstract}

\maketitle

\section{Introduction}

Magnetic skyrmions are textures of non-trivial spins topologically stable. Because of this stability, they can be manipulated with very low currents and have been proposed for use in new technologies, such as in data storage \cite{SCR2013,TOM2014}, logic devices \cite{app3}, 2D skyrmions crystals \cite{Gilbert_D}, and several types of purely solid-state spin-based information processing devices \cite{app1,app2}. The nucleation and stabilization of these textures \cite{app1,sky_1,sky_2,sky_3,sky_5,sky_6} is due to the existence of an antisymmetric interaction known as Dzyaloshinskii-Moriya interaction (DMI) \cite{DMI_1,DMI_2,DMI_3}, which appears in certain materials, whose origin is due to two possible causes: a strong spin-orbit coupling (SOC) and/or a lack in inversion symmetry \cite{DMI_1,DMI_2}. 

The bulk DMI is due to a crystallographic ordering of the atoms, which gives rise to the two causes mentioned above. Bloch domain walls are favoured when we are in the presence of the bulk DMI, so Bloch skyrmions (BS) are stabilized \cite{sky_2,BS_2,BS_3,BS_4}. Materials with this type of interaction are generally referred to as chiral magnets and in particular, skyrmions can be recognized as a chiral vortex-like structure. On the other hand, the interfacial DMI arises from a breaking of inversion symmetry on the surface, due to a SOC that occurs with an adjacent film \cite{app1}. This interaction allows stabilizing the so-called N\'eel skyrmion (NS) \cite{SCR2013,sky_6,NS_1}, also known as the hedgehog-like configuration, which is characterized by a non-zero contribution in the radial direction. It is important to note that NS has a comparative advantage over BS that would allow its use in the development of new technologies. This advantage arises because NS have been stabilized even at room temperature \cite{RT1,RT2,RT3}, while BS are only stable over a short temperature range \cite{BSRT}.

Recently, interest has also been devoted to isolated skyrmions confined in magnetic dots \cite{SCR2013,sky_6,RT1,RT2,RT3,G2015,SA2016,Dip_1}. In this geometry, both interfacial DMI and magnetic anisotropy are required to stabilize a NS \cite{SCR2013,sky_6,Dip_1}, whereas BS can be stabilized in the absence of DMI , as long as there is a magnetic anisotropy \cite{G2015,SA2016,DAI2013}. Previous studies that investigated skyrmions on cylindrical geometries show that both DMI and anisotropy play a fundamental role in stabilizing skyrmions \cite{SCR2013,sky_6,G2015,SA2016,Dip_1}. However, many of these studies neglected the dipolar interaction, as is usually done for thin films, where its contribution is quite small. Now, this approach must be treated with care, since considering or not the dipolar interaction modifies the magnetic phase diagram that considers the skyrmion configuration in nanodots \cite{Dip_1}.

To date, there are few analytical studies that investigate the stability of NS in cylindrical systems \cite{sky_6,NS2}, but none of them considered the dipolar interaction. Thus, in this paper we have proposed a general ansatz for the magnetization that would describe a confined NS in a nanodot, considering exchange, uniaxial anisotropy, Dzyaloshinskii-Moriya, and dipolar contributions, and analyse its stability in terms of its magnetic parameters. Finally, we will show that considering or not the dipolar interaction produces a variation in the radius of the skyrmion. 

\section{Analytical Description}

\begin{figure}[ht]
\begin{center}
\includegraphics[width=10cm]{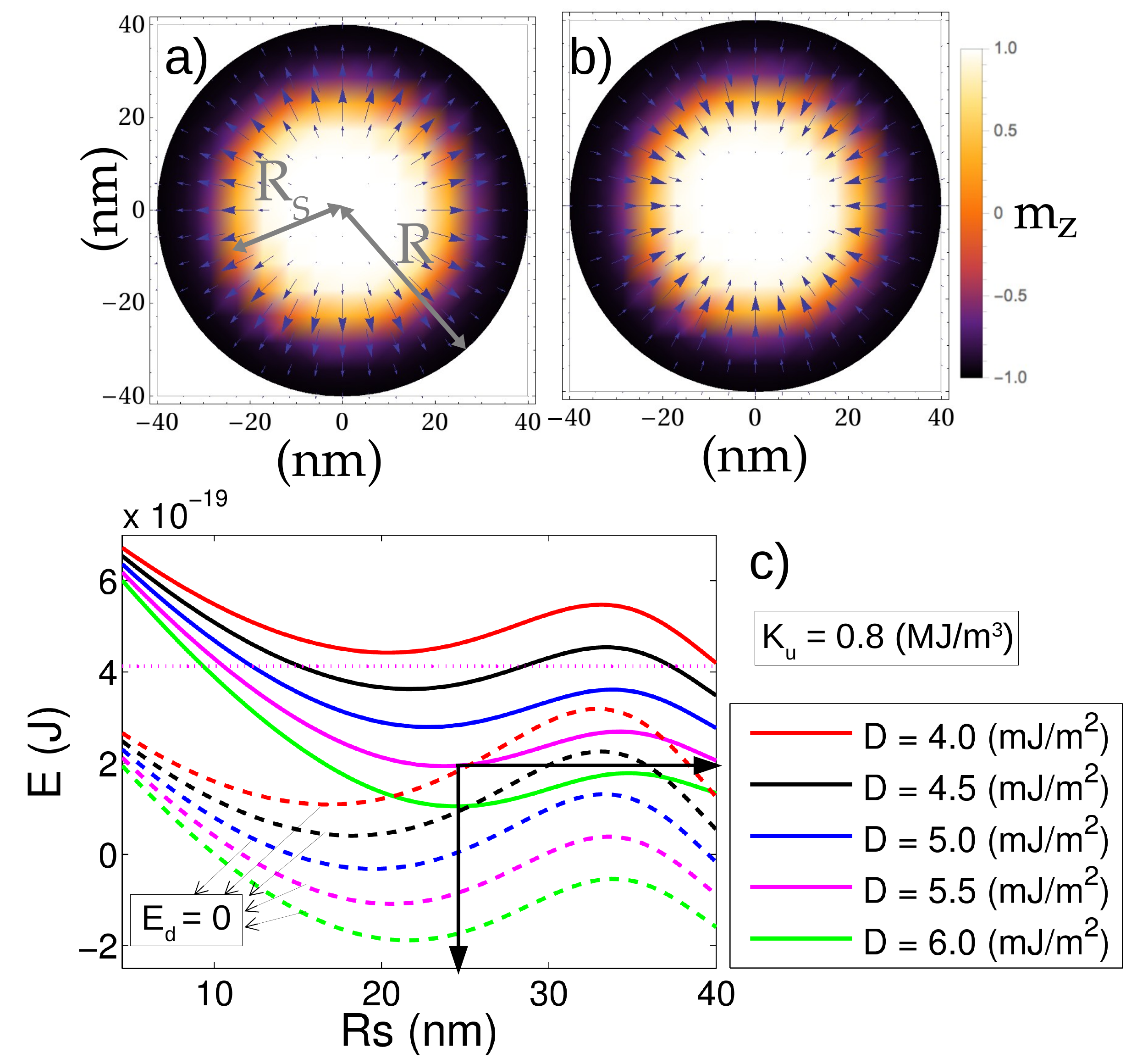}
\end{center}
\caption{(Color online) Profile of the magnetization of a NS in a 40 nm radio nanodot obtained from the analytical model (with $n = 10$ and $R_s = 25$ nm) for different chirality $C = +1$ (a) and $-1$ (b). In these figures $m_z(r)$ is presented as a color density plot, while the $r$-component vector is presented with arrows. (c) Energy of a NS as a function of $R_s$ for $C = 1$, an anisotropy constant $K_u = 0.8 \, (MJ/m^3)$ and different values for the Dzyaloshinskii-Moriya constant $D$. The solid lines consider the dipolar energy, while the dashed lines do not. The red dotted line represents the out-of-plane ferromagnetic configuration.}
\label{skyrmion_fig}
\end{figure}

We are interested in investigating a NS in an ultra-thin dot of radius $R$ and thickness $L$. The magnetization of this NS can be written as $\vec{m} = m_z(r)\hat{z} + m_r(r) \hat{r} $, where $\hat{z}$ and $\hat{r}$ are unitary vectors in cylindrical coordinates (we have considered the $z$-direction along the axis of symmetry of the dot). In particular, we have proposed a general expression for the $z$ component of the magnetization of NS in the form: 
\begin{equation}
\label{mz_n}
m_z(r) = \frac{1-(r/R_s)^n}{1 + (r/R_s)^n}
\end{equation}
where $R_s$ corresponds to the radius of the skyrmion, i.e., it is the value of $r$ for which $m_z$ vanishes; and $n$ is an even positive integer, $n = 2, 4, 6, 8, \cdots$. Since $\vec{m}$ is a unitary vector, the $r$-component is given by $m_r(r) = C\sqrt{1 - m_z^2(r)}$, with $C = \pm 1$ (the chirality of skyrmion). As an illustration, Fig. \ref{skyrmion_fig} shows the magnetization of the NS obtained from the theoretical model presented here (for $n = 10$) when $C = +1$ (a) and $-1$ (b). Besides, in order to confirm that our model, Eq. \eqref{mz_n}, effectively describes a skyrmion, we have calculated the topological skyrmion number $S = \frac{1}{4 \pi} \int dxdy \, \vec{m} \cdot (\frac{\partial \vec{m}}{\partial x} \times \frac{\partial \vec{m}}{\partial y} )$, which should be a number close to 1. Defining $\delta= R_s/R$ we have $S = 1/(1 + \delta^n) \approx 1 $, since for a NS it satisfies $0 < R_s < R$. 

It is interesting to mention that the $z$-component of the proposed magnetization in Eq.\eqref{mz_n} is a generalization of the particular case $n = 2$ previously used by Guslienko \cite{G2015} for the study of BS in a nanodot with perpendicular anisotropy in the absence of external magnetic field and DMI. In this article the author points out that skyrmions can be stabilized even in the absence of DMI, however, for NS this interaction is responsible for defining the chirality of the skyrmion \cite{app2,SCR2013}, and therefore must be considered. In addition to DMI, we will also consider exchange energy, uniaxial anisotropy, and dipole energy, the latter not considered in previous studies \cite{sky_6,NS2}.

The exchange energy is given $E_\text{ex}= A \int_V dV \sum_{i=x,y,z}(\vec{\nabla} m_i)^2$, where $A$ is the stiffness constant. Solving for the profile of the magnetization proposed in this article, Eq. (1), we obtain
\begin{equation}
\label{Eex}
E_\text{ex} =  2 \pi A L\frac{4 + n^2}{n(1+\delta^n)} ,
\end{equation}
where $L$ is the thickness of the nanodot.

The uniaxial anisotropy energy is given by the expression $E_\text{u} = K_u \int_V dV (1-mz^2)$, where $K_u$ is the uniaxial anisotropy constant and $V= \pi R^2 L$ is the volume of the dot. Solving for NS we have
\begin{eqnarray}
\label{Eu}
E_\text{u} &=& K_u V  - \frac{ K_u V}{n(1+\delta^{-n})} \left\lbrace n\, (1+\delta^{-n}) + 8\right. \nonumber \\
&& \left. -8\, (1+\delta^{-n})\, _2\text{F}_1(1;2/n;1+2/n;-\delta^{-n}) \right\rbrace,
\end{eqnarray}
where $_2\text{F}_1(a;b;c;z)$ is a hypergeometric function.

The Dzyaloshinskii-Moriya interaction depends on the size of the material and its crystallographic symmetry. For the ultra-thin magnetic nanodots considered in this article, the DMI is given by the surface integral $E_\text{DM} = -LD \int_S dS [m_z \vec{\nabla} \cdot\vec{m} - (\vec{m} \cdot \vec{\nabla})m_z]$ \cite{sky_6,DM_NS}, being $D$ the Dzyaloshinskii-Moriya constant. Solving this integral we obtain
\begin{eqnarray}
\label{EDM}
&&E_\text{DM} = -2  C\,D \, \frac{V}{R} \, \frac{\delta^{-n/2}}{n} \nonumber \\
&& \times \left\lbrace \frac{4}{1+\delta^{-n}} -\frac{\delta^{-n}}{3n+2}(n^2-4) \, _2\text{F}_1(1;\frac{3}{2}\hspace{-0.1cm}+\hspace{-0.1cm}\frac{1}{n};\frac{5}{2}\hspace{-0.1cm}+\hspace{-0.1cm}\frac{1}{n};-\delta^{-n})  \right. \nonumber\\
&& \left. + \, (n-2)[ \, 1 + \, _2\text{F}_1(1;\frac{1}{2}\hspace{-0.1cm}+\hspace{-0.1cm}\frac{1}{n};\frac{3}{2}\hspace{-0.1cm}+\hspace{-0.1cm}\frac{1}{n};-\delta^{-n})] \right\rbrace .
\end{eqnarray}

The last contribution to energy is due to the dipolar interaction, which is generally neglected due to the consideration of ultra-thin films. However, Beg \textit{et al.} \cite{Dip_1} showed that this assumption is not always correct, and should be considered. The dipolar contribution is given by $E_\text{d} = \mu_0 M_0/2 \int_V dV \vec{m}\cdot \vec{\nabla}U(\vec{r})$, where $M_0$ is the saturation magnetization and $U(\vec{r})$ is the magnetostatic potential \cite{Aharoni1996}. After calculating the magnetostatic potential in cylindrical coordinates, and some algebraical manipulation, the dipolar energy for a NS is given by
\begin{eqnarray}
\label{Ed}
E_\text{d} = \mu_0 \pi M_0^2 && \int_0^\infty  dk \left\lbrace \, g_{0,z}^2(k) \, [1- e^{-kL}] + \,  g_{1,r}^2(k) \, [kL + e^{-kL}-1] \right\rbrace,
%E_\text{d} = \mu_0 \pi M_0^2 && \int_0^\infty  dk \left\lbrace \, g_{0,z}^2(k) \, [1- e^{-kL}] \right. \nonumber \\
%&&\left. + \,  g_{1,r}^2(k) \, [kL + e^{-kL}-1] \right\rbrace,
\end{eqnarray}
where we have defined the functions
\begin{eqnarray}
g_{\nu,z}(k) &=& \int_0^R dx \, x \, J_\nu(kx) \, m_z(x) \\
g_{\nu,r}(k) &=& \int_0^R dx \, x \, J_\nu(kx) \, m_r(x).
\end{eqnarray}
Here $J_\nu(z)$ corresponds to the first kind Bessel function. Finally, the total energy $E$ for the NS is given by the sum of Eqs \eqref{Eex}, \eqref{Eu}, \eqref{EDM} and \eqref{Ed}.

\section{Example of model application: ultra-thin Co/Pt nanodots}

In the framework of this article we have obtained an analytical expression for the energy $E$ of a NS whose profile of magnetization is given by Eq. (1). The idea is that the radius of the skyrmion $R_s$ is the minimizable parameter. In order to verify that the proposed magnetization profile is adequate and correctly describe a NS, as well as the analytical expressions obtained for the different energy contributions, we will obtain analytical results for ultra-thin Co/Pt nanodots, with the aim of comparing these results with numerical simulations previously reported \cite{SCR2013}. In this way, we have considered the geometric and magnetic parameters used in this reference, that is, $R = 40 \,(nm)$, $L = 0.4 \,(nm)$, $A = 15 \, (pJ/m)$ and $M_0 = 580 \,(kA/m)$, while the magnetic parameters $D$ and $Ku$ were left as free parameters. In addition, as shown by Sampaio \textit{et al.} \cite{SCR2013}, the interaction of Co films on Pt generates a strong spin-orbit interaction, which can be described by the superficial DMI considered in this article. At this point it is important to consider that the chirality of the skyrmion is given by Eq. (4), where the term $CD$ appears. In this way, for $D> 0$ the skyrmion with $C=1$ will always exhibit lower energy than the skyrmion with chirality $C = -1$. Therefore, from now on  we will always consider $C = 1$.

As a first step we are interested in determining a reliable $n$ value for the model proposed in Eq. (1), which allows to adequately describe the profile of the magnetization of a NS, giving values for the radius of skyrmion, $D$ and $K_u$ comparable to those obtained previously by Sampaio \textit{et al.} \cite{SCR2013}. For this, we have carry out numerical simulations very similar as the one performed in \cite{SCR2013} in order to compare with the analytical results for different $n$-values (the details of the analysis can be found in the appendix). For the range of parameters considered, we find that $n = 2$ does not correctly describe a NS because the energy of this skyrmion is always higher than the energy of the out-of-plane ferromagnetic configuration (F), as shown in Figs. 5 and 6 of the appendix. However, from the same figures we can conclude that $n = 10$ is a reliable parameter to describe this type of skyrmions for the range of parameters considered. Therefore, from now on we will use $n = 10$.

Figure 1(c) shows the energy of a NS (for $n = 10$, $C=1$ and $K_u = 0.8 \, (MJ/m^3)$) as a function of its radius $R_s$ for $D = 4.0, 4.5, 5.0, 5.5$ and $6.0 \, (mJ/m^2)$ (red, black, blue, magenta and green curves, respectively). The solid lines consider the dipolar energy, while the dashed lines do not. From this figure we can conclude that the dipolar term given by the Eq. (5) cannot be neglected, as it happened in previous studies \cite{sky_6,NS2}, and must be considered for the calculation of the energy of a NS. In this figure we have also plotted the energy of the out-of-plane ferromagnetic configuration (red dotted line). This allows to visualize that $D = 4.0$ (red solid line) is not enough to obtain a NS in the nanodot (for this particular $Ku = 0.8 \, (MJ/m^3)$). For the rest of the curves we have obtained a NS, whose energy and radius were obtained by minimizing energy $E$ as a function of $R_s$. It is interesting to note that the N\'eel skyrmions are stable only for $D > 5.0 \, (mJ/m^2)$ (as denoted by the arrow) and are unstable for $4.0 < D < 5.0 \, (mJ/m^2)$.

Figure 2 shows the energy $E$ (a) and radius of the skyrmion $R_s$ (b) as a function of $D$, for $Ku = 0.8 \, (MJ/m^3)$. The red hollow symbols correspond to the analytical results obtained for $n = 10$, while the blue solid symbols correspond to micromagnetic simulations. In particular, for Fig. 2(a) we have that the squares correspond to the out-of-plane ferromagnetic configuration while the circles correspond to a NS. From this figure we can see that the energies both for the analytical calculations and those obtained by numerical simulations show a similar behaviour and are in the same order of magnitude. From the numerical simulation we can obtain, for $Ku = 0.8 \, (MJ/m^3)$, a NS from $D = 4 \, mJ/m^2$ onwards, while from the theoretical model we obtain a NS just from $D \approx 4.3 \, (mJ/m^2)$ (a value similar to the numerical one), but as we mentioned above, the skyrmion configuration is only stable for $D > 5.0 \, (mJ/m^2)$. It is important to note that for $D > 6 \, (mJ/m^ 2)$ the simulations relax to a configuration with multiple domains. On the other hand, in Fig. 2(b) we can observe an excellent agreement between the radius of the skyrmion obtained by means of micromagnetic simulations and analytical calculations, validating in some way the theoretical model proposed for the profile of the magnetization of the NS. In particular we observe that the radius of the skyrmion increases slightly as the value of D increases.

\begin{figure}[h]
\begin{center}
\includegraphics[width=10cm]{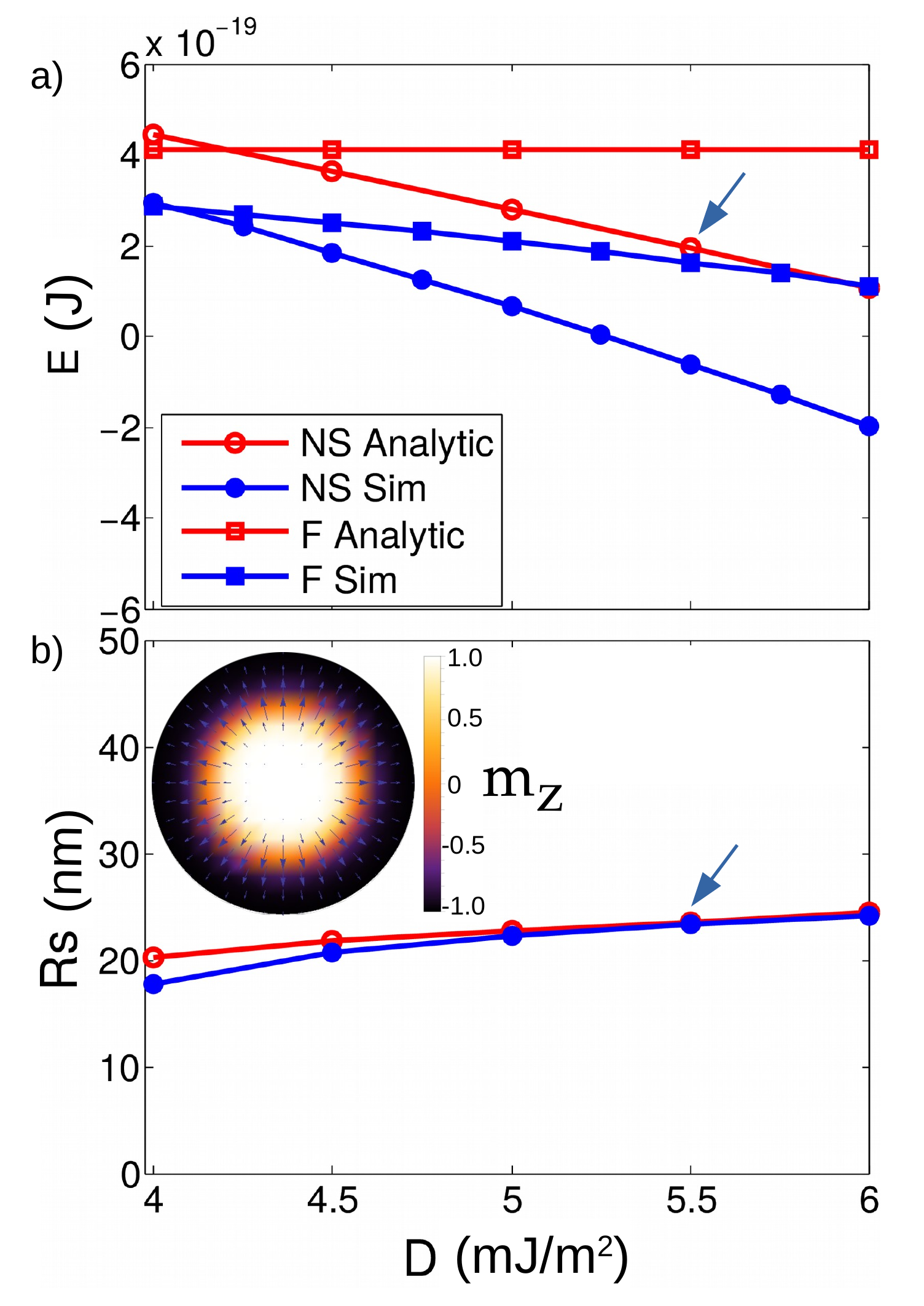}
\end{center}
\caption{(color online) Energy $E$ (a) and radius of the skyrmion $R_s$ (b) as a function of $D$, for $Ku = 0.8 \, (MJ/m^3)$. The red hollow symbols correspond to the analytical results obtained for $n = 10$, while the blue solid symbols correspond to micromagnetic simulations \cite{SCR2013}. The profile of the magnetization shown as inserted in figure (b) corresponds to the configuration that is highlighted with arrows.}
\label{Esk_D}
\end{figure}

Figure 3 shows the energy $E$ (a) and radius of the skyrmion $R_s$ (b) as a function of $Ku$, for $D = 5.5 \, (mJ/m^2)$. The red hollow symbols correspond to the analytical results obtained for $n = 10$, while the blue solid symbols correspond to micromagnetic simulations. In particular, in Fig. 3(a) both the analytical calculations and numerical simulations show a good agreement. From the numerical simulation we can obtain, for $D = 5.5 \, (mJ/m^2)$, a NS for values below $Ku= 1.4 \, (MJ/m^3)$, while from the theoretical model we obtain a NS for values below $Ku=1.3 \, (MJ/m^3)$. It is important to note that for $K_u < 0.4 \, (MJ/m^3)$ the simulations relax to a configuration with multiple domains. On the other hand, in Fig. 3(b) we can observe that, although the values obtained by numerical simulations and analytical calculations for the skyrmions radius are in the order of magnitude, it is also true that the adjustment is worse than the obtained in Fig. 2(b). Moreover, from the theoretical model we obtain that the radius of the skyrmion decreases as we increase the uniaxial anisotropy constant.

\begin{figure}[h]
\begin{center}
\includegraphics[width=10cm]{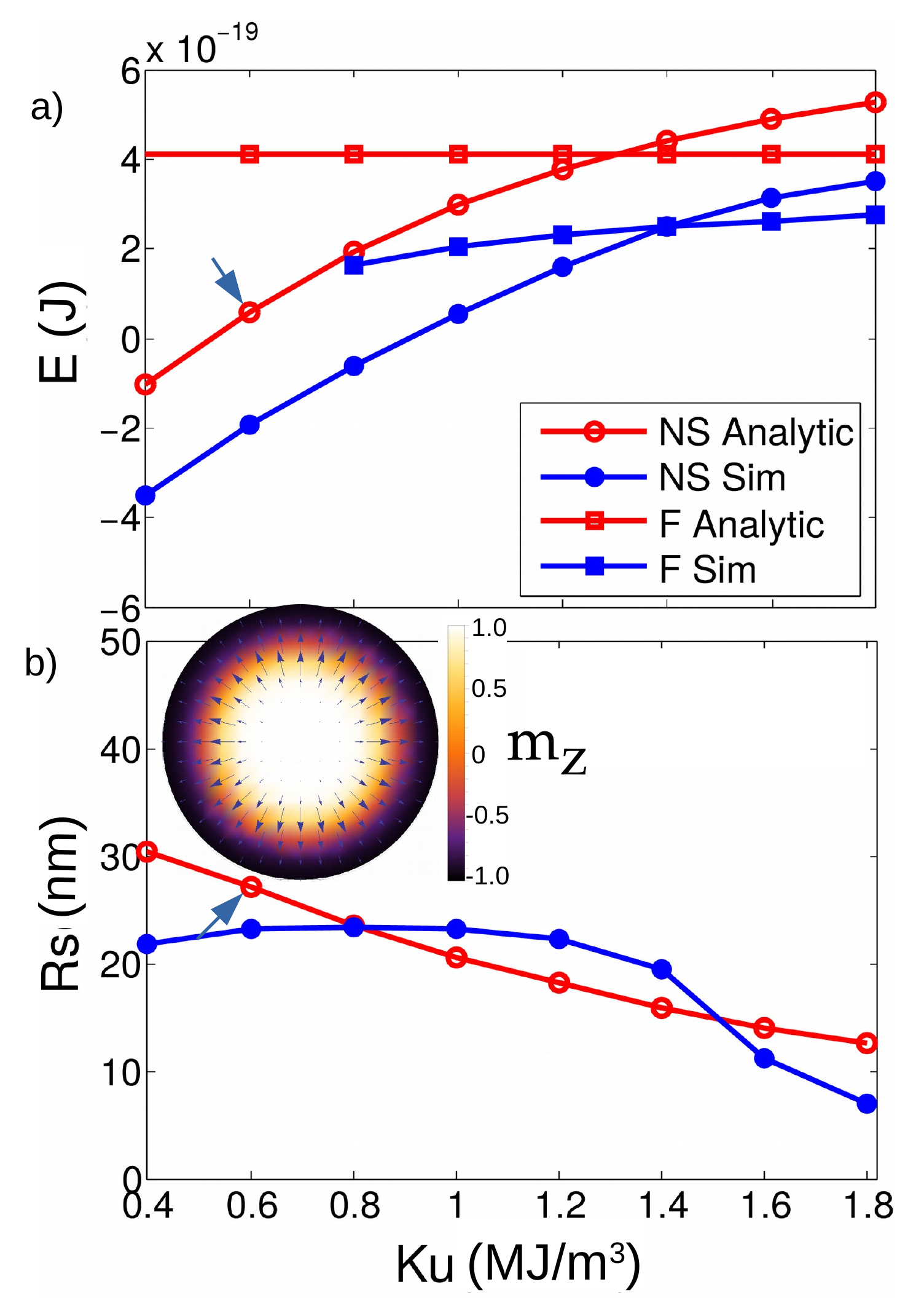}
\end{center}
\caption{(color online) Energy $E$ (a) and radius of the skyrmion $R_s$ (b) as a function of $Ku$, for $D = 5.5 \, (mJ/m^2)$. The red hollow symbols correspond to the analytical results obtained for $n = 10$, while the blue solid symbols correspond to micromagnetic simulations \cite{SCR2013}. The profile of the magnetization shown as inserted in figure (b) corresponds to the configuration that is highlighted with arrows.}
\label{Esk_Ku}
\end{figure}

\section{Conclusions}

In conclusion, we have proposed for the N\'eel skyrmion a general expression for the component of magnetization perpendicular to the dot, given by $m_z(r) = [1-(r/R_s)^n]/[1 + (r/R_s)^n]$. This expression is a generalization of the particular case $n = 2$ previously used by Guslienko \cite{G2015} for the study of BS in a nanodot with perpendicular anisotropy. In particular, we have obtained an analytical expression for the energy of N\'eel skyrmions in ultra-thin nanodots considering exchange, uniaxial anisotropy, Dzyaloshinskii-Moriya, and dipolar contributions, the latter not considered in previous studies \cite{sky_6,NS2}.

 As proof of concept, we have calculated the energy of a N\'eel skyrmion in an ultra-thin Co/Pt cylinder, and we found that the dipolar contribution cannot be neglected and that both Dzyaloshinskii-Moriya interaction and anisotropy play an important role to stabilize the skyrmion. We have obtained a good agreement between our analytical calculations and previously published micromagnetic simulations \cite{SCR2013} for $n = 10$. In particular, we have obtained that for a Dzyaloshinski Moriya constant $D = 5.5 \, (mJ/m^2)$, it is possible to stabilize a N\'eel skyrmion for $K_u$ in the range, $0.4 \, (MJ/m^3)< K_u <1.3 \, (MJ/m^3)$, whereas for $K_u = 0.8 \, (MJ/m^3)$, the skyrmion stabilizes for $5.0 \, (mJ/m^2) < D <6.0 \,  (mJ/m^2) $. Finally, the proposed theoretical model (which considers the dipolar interaction) makes it possible to predict the radius of the skyrmion quite accurately. In this way, this model can be used to predict ranges of stability of N\'eel skyrmions, for different magnetic and geometric parameters, which have been proposed for use in spintronic devices.

\section*{Acknowledgments}

We thank Felipe Tejo and Sebastian Allende for their insightful comments. This work was supported by Fondecyt Grant 1150952, DICYT Grant 041631EM-POSTDOC from VRIDEI-USACH, Financiamiento Basal para Centros Cient\'{\i}ficos y Tecnol\'{o}gicos de Excelencia FB0807, and Conicyt-PCHA/Doctorado Nacional/2014.

\section*{Appendix}

In order to obtain a reliable value of $n$ for our model, in Fig. 4 we compare the energy $E$ of a NS confined in an ultra-thin Co/Pt nanodot as a function of the radius of the skyrmion $R_s$ for different values of $n$. The left column (Figs. 4(a), (b) and (c)) shows the total energy for different values of $D$ (with $Ku=0.8(MJ/m^3)$), while the right column (Figs. 4(d), (e) and (f)) shows the total energy for different values of $Ku$ (with $D = 5.5 \, (mJ/m^2)$). The red dashed horizontal line represents the energy of the out-of-plane ferromagnetic configuration for all images. From the Fig. 4 we can conclude that the stability of the N\'eel skyrmions depends on the geometric and magnetic parameters considered. 

For example, Fig. 4(a) shows that for $D=4.0 \, (mJ/m^2)$ and $Ku=0.8 \, (MJ/m^3)$, it is not possible to stabilize a NS, since out-of-plane ferromagnetic configuration always exhibits a lower energy, regardless of the value of $n$. However, by increasing $D$ (see Figs. 4(b) and (c)) we find that the N\'eel skyrmions are energetically more favourable than the out-of-plane ferromagnetic state. It is interesting to note that the energy of the NS exhibits a non-monotonic behaviour with $n$, decreasing energy as we increase $n$ to a certain value, after which it begins to increase (see for example the case of $n = 18$). Analogously, Fig. 4(f) shows that for $D = 5.5 \, (mJ/m^2)$ and $Ku=1.4 \, (MJ/m^3)$, it is not possible to stabilize a NS, since out-of-plane ferromagnetic configuration always exhibits a lower energy, regardless of the value of $n$. However, by decreasing $Ku$ (see Figs. 4(d) and (e)) we find that the N\'eel skyrmions are energetically more favourable than the out-of-plane ferromagnetic state.

\begin{figure}[h]
\begin{center}
\includegraphics[width=10cm]{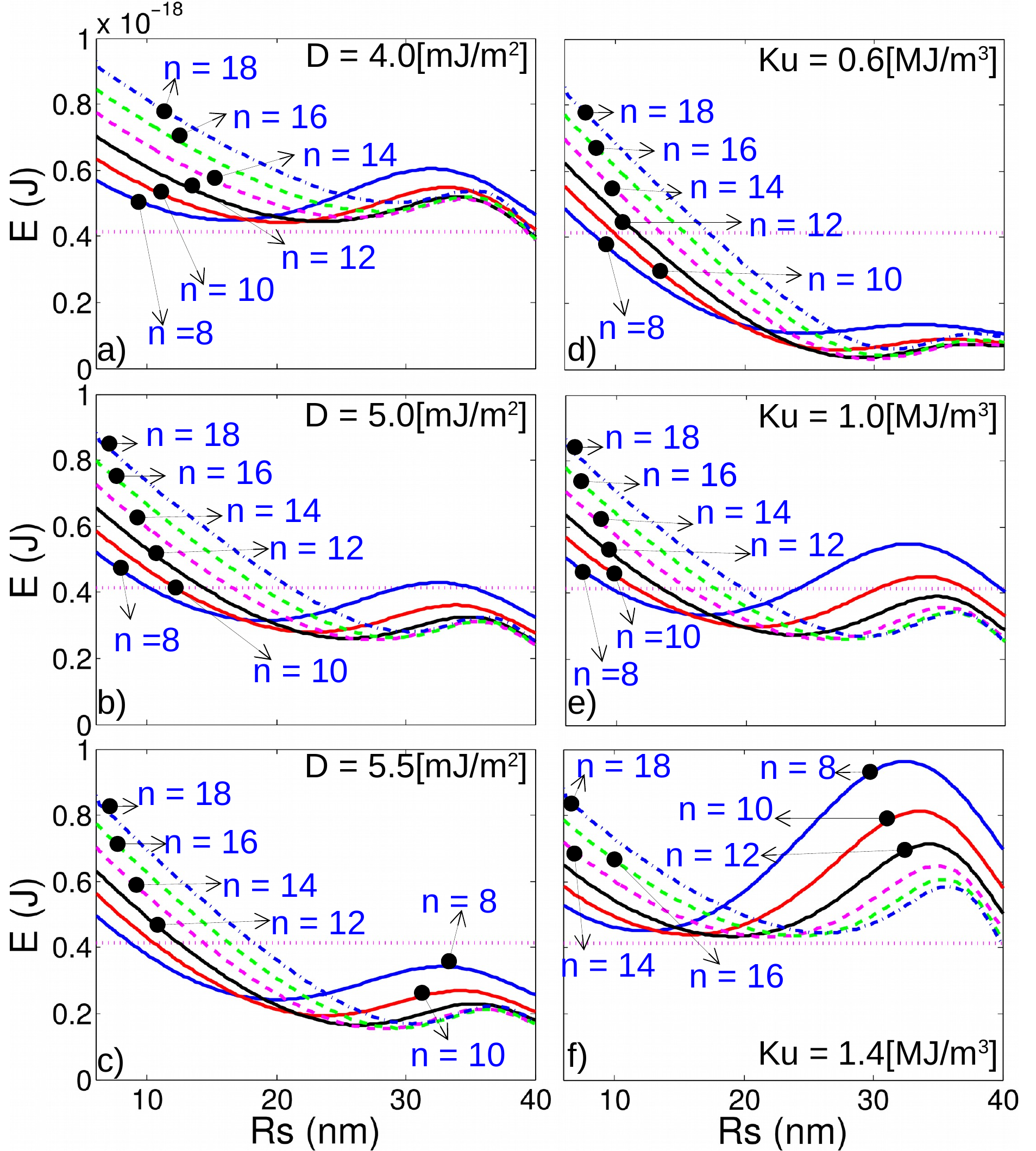}
\end{center}
\caption{(color online) Energy $E$ of a NS as a function of its radius $R_s$ for different values of $n$. The figures in the left column consider $Ku = 0.8 \, (MJ/m^3)$ for different values of $D$, while the figures in the right column consider $D = 5.5 \, (mJ/m^2)$ for different values of $Ku$.}
\label{Esk_n}
\end{figure}

So far we have concluded that the energy of the skyrmion follows a non-monotonic behaviour with the parameter $n$, but without yet defining a reliable value. In this way, our analysis continues with the analytical calculation of the energy $E$ and radius of the skyrmion which minimizes its energy $R_s$ as a function of the parameters $D$ (see Fig. 5) and $Ku$ (see Fig. 6) for different values of $n$. The idea is to compare these analytical results with micromagnetic simulations \cite{SCR2013} and obtain a reliable $n$ value for the proposed model. From Figs. 5 and 6 we can see that for $n> 8$ the analytical model reproduces quite well the behaviour obtained from the micromagnetic simulations (except for very large $n$ values). In particular, we find that $n = 10$ is the one that best replicates the behaviour of the radius of the skyrmion $R_s$ as a function of the $D$ and $Ku$ parameters (see in particular Fig. 5b).

\begin{figure}[h]
\begin{center}
\includegraphics[width=10cm]{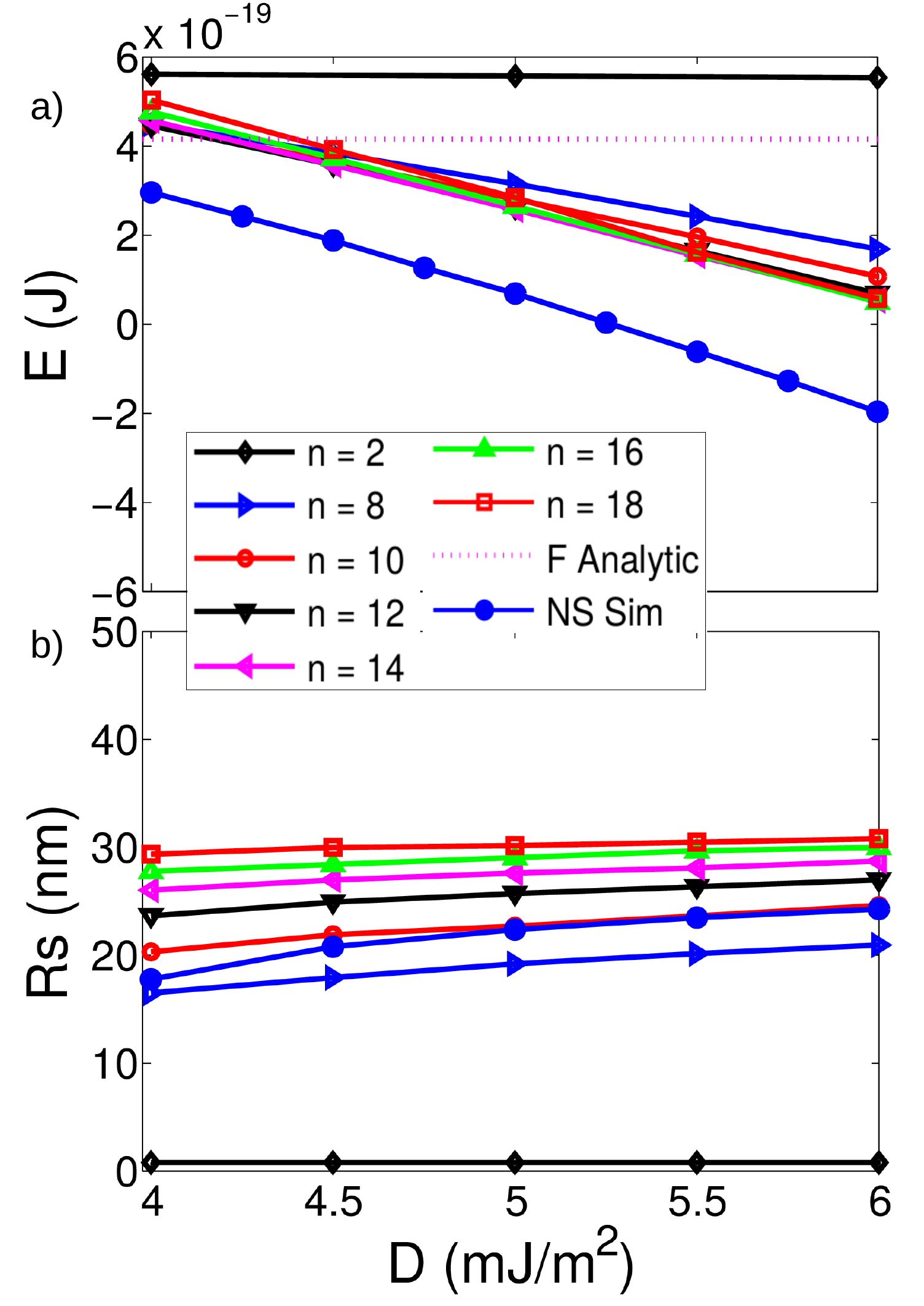}
\end{center}
\caption{(color online) Energy $E$ (a) and radius of the skyrmion $R_s$ (b) for different $n$-values as a function of $D$, for $Ku = 0.8 \, (MJ/m^3)$. The blue solid symbols correspond to micromagnetic simulations \cite{SCR2013}.}
\label{Esk_D_n}
\end{figure}

In summary, for the range of geometric and magnetic parameters investigated, the model works correctly for $8<n<18$, and we find that $n = 10$ is a reliable parameter for the model. In addition, we have compared the profile of the magnetization of the NS obtained analytically with that of the numerical simulation (not shown here), and we have come to the same conclusion.

\begin{figure}[h]
\begin{center}
\includegraphics[width=10cm]{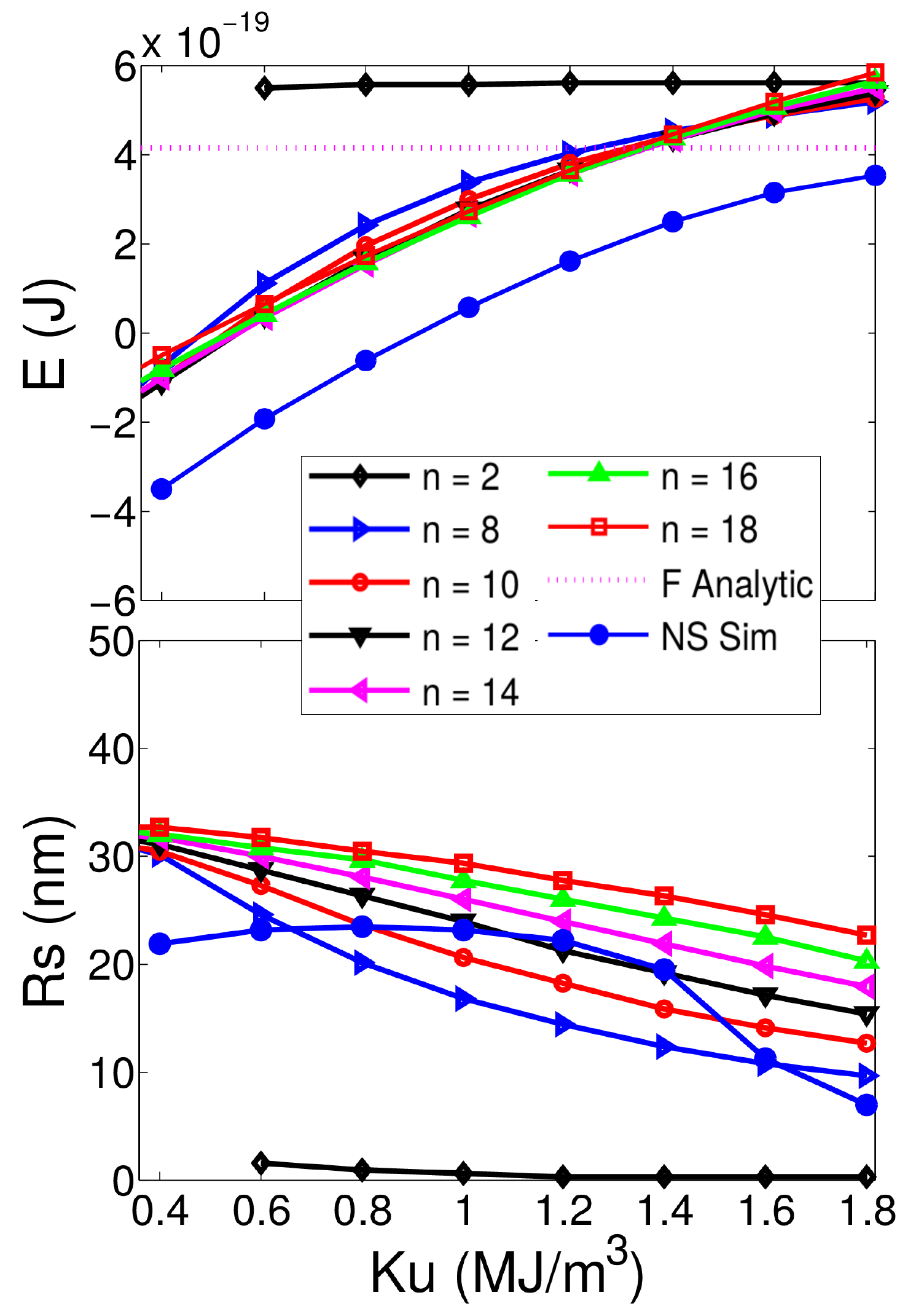}
\end{center}
\caption{(color online) Energy $E$ (a) and radius of the skyrmion $R_s$ (b) for different $n$-values as a function of $Ku$, for $D = 5.5 \, (mJ/m^2)$. The blue solid symbols correspond to micromagnetic simulations \cite{SCR2013}.}
\label{Esk_Ku_n}
\end{figure}

\end{document}